\documentclass[reprint,showpacs,preprintnumbers,amsmath,amssymb, citeautoscript,prl,superscriptaddress]{revtex4-1}
\usepackage{graphicx,wasysym}% Include figure files
\usepackage{dcolumn}% Align table columns on decimal point
\usepackage{bm}% bold math
\setcitestyle{super}

\usepackage{tabulary}
\usepackage{hyperref}

\newcommand{\tN}{t_{\mathrm{N}}}

\newcommand{\sigmaNE}{\sigma_{\mathrm{id}}}

\newcommand{\kB}{k_\mathrm{B}}

\newcommand{\ve}[1]{\mathbf{#1}}
\usepackage{bbold}

\usepackage{xcolor}

\begin{document}

\preprint{1}

\title{Field-dependent ionic conductivities from generalized fluctuation-dissipation relations}
\author{Dominika Lesnicki}
\affiliation{Sorbonne Universit\'e, CNRS, Physico-Chimie des \'electrolytes et
Nanosyst\`emes Interfaciaux, F-75005 Paris, France}
\author{Chloe Y. Gao}
\affiliation{Department of Chemistry, University of California, Berkeley, California}
\author{Benjamin Rotenberg}
\affiliation{Sorbonne Universit\'e, CNRS, Physico-Chimie des \'electrolytes et
Nanosyst\`emes Interfaciaux, F-75005 Paris, France}
\affiliation{R\'eseau sur le Stockage Electrochimique de l'Energie
(RS2E), FR CNRS 3459, France}
\author{David T. Limmer}
 \email{dlimmer@berkeley.edu}
 \affiliation{Department of Chemistry, University of California, Berkeley, California}
\affiliation{Kavli Energy NanoScience Institute, Berkeley, California}
\affiliation{Materials Science Division, Lawrence Berkeley National Laboratory, Berkeley, California}
\affiliation{Chemical Science Division, Lawrence Berkeley National Laboratory, Berkeley, California}

\date{\today}
\begin{abstract}
We derive a relationship for the electric field dependent ionic conductivity in terms of fluctuations of time integrated microscopic variables. We demonstrate this formalism with molecular dynamics simulations of solutions of differing ionic strength with implicit solvent conditions and molten salts. These calculations are aided by a novel nonequilibrium statistical reweighting scheme that allows for the conductivity to be computed as a continuous function of the applied field. In strong electrolytes, we find the fluctuations of the ionic current are Gaussian and subsequently the conductivity is constant with applied field. In weaker electrolytes and molten salts, we find the fluctuations of the ionic current are strongly non-Gaussian and the conductivity increases with applied field. This nonlinear behavior, known phenomenologically for dilute electrolytes as the Onsager-Wien effect, is general and results from the suppression of ionic correlations at large applied fields, as we elucidate through both dynamic and static correlations within nonequilibrium steady-states.   
\end{abstract}

\maketitle

Advances in the fabrication of nanofluidic devices have enabled the study of
transport processes on small scales, where novel phenomena emerge from the
interplay of confinement, fluctuations and molecular
granularity.\cite{esfandiar2017size,mouterde2019molecular,siria2013giant,secchi2016scaling}
Some of the most striking recent observations have been in electrokinetic
transport of electrolyte solutions confined to nanometer dimensions. In such
systems, large thermodynamic gradients can be generated, driving nonlinear
responses such as Coulomb blockade and current rectification.\cite{kim2007concentration,karnik2007rectification,guan2011field,vermesh2009fast,feng2016observation} 
At the same time, and independently, theoretical developments have permitted the
application of response theory to systems far from equilibrium. In particular,
generalized fluctuation-dissipation theorems have been formulated for linear
responses of nonequilibrium steady states, or nonlinear responses around
equilibrium
states.\cite{speck2006restoring,prost2009generalized,baiesi2009fluctuations,baiesi2013update,nemoto2011thermodynamic,gaspard2013multivariate,gao2018nonlinear}
These coincident developments provide a way for predicting transport
relationships from molecular properties, and using such relationships to design
nanofluidic devices. Building on these previous work, and inspired by emerging
challenges in nanofluidic devices, we develop a theory and accompanying
numerical technique to efficiently compute the electric field dependent
conductivity in ionic solutions. We recover behavior similar to the
Onsager-Wien effect in the dilute limit,\cite{onsager1957wien} however our
calculations are valid across all concentration regimes, and arbitrary nonlinear
responses.  Our approach is general and can be extended to other systems or transport processes where a connection between nonlinear transport behavior and underlying microscopic dynamics is desired.

We consider a system of $N$ ions composed of $N_a$ anions and $N_c$ cations, in a volume $V$ and fixed temperature, $T$.
%, in implicit solvent conditions
The ions' positions and velocities are denoted, $\ve{r}^N=\{ \ve{r}_1,\ve{r}_2, \dots,\ve{r}_N \}$ and $\ve{v}^N=\{\ve{v}_1,\ve{v}_2, \dots,\ve{v}_N\}$, respectively. These variables evolve according to an underdamped Langevin equation,
\begin{equation}
\label{Eq:Lang}
\dot{\ve{x}}_i=\ve{v}_i \, ,\quad m_i \dot{\ve{v}}_i = - \zeta_i \ve{v}_i + \ve{F}_i \left (\ve{r}^N \right) + z_i \ve{E}  + \bm{\eta}_i
\end{equation}
where $m_i$ and $z_i$ are the $i$th particle's mass and charge, $\zeta_i$ is the
friction from the implicit solvent, and $\ve{F}_i(\ve{r}^N)$ is the
interparticle force on ion $i$, which we take as a pairwise sum of electrostatic interaction with dielectric constant $\epsilon_s$ and non-electrostatic (short-range repulsion and long-range dispersion) forces. Further details of the force fields can be found in the Supplemental Material (SM).
%screened Coulomb interactions with dielectric constant $\epsilon_s$, and Lennard-Jones forces. 
Each cartesian component of the random force, $\ve{\eta}_{i\alpha}$, obeys
Gaussian statistics with mean $\langle \ve{\eta}_{i\alpha} \rangle =0$ and
variance $\langle \ve{\eta}_{i\alpha}(t)\ve{\eta}_{j\beta}(t') \rangle = 2 \kB
T \zeta_i \delta_{ij} \delta_{\alpha\beta} \delta(t-t')$, where $\kB$ is
Boltzmann's constant. Finally, $\ve{E}$ denotes an applied electric field, with magnitude $E$, which drives an ionic current through the periodically replicated system. This equation of motion does not conserve momentum, and thus hydrodynamic effects are explicitly neglected throughout. %\emph{More?}

To compute the ionic conductivity as a function of electric field, we aim to relate dynamic quantities of the system at a reference field, to those of a system perturbed by an additional applied field. 
Given the equation of motion in Eq.~\ref{Eq:Lang}, the probability of observing a trajectory, $\ve{X}(\tN)$, or sequence of positions and velocities over an observation time, $\tN$,  with an applied field, is
\begin{equation}
\label{Eq:Measures}
P_{\ve{E}}[\ve{X}(\tN)] \propto e^{-\beta U_{\ve{E}}[\ve{X}(\tN)]}
\end{equation}
where for the uncorrelated Gaussian noise, we have an Onsager-Machlup stochastic action\cite{cugliandolo2019building} of the form
\begin{align}
U_{\ve{E}}[\ve{X}(\tN)]& =
\sum_{i=1}^N  \int_0^{\tN}  dt \frac{\left [m_i \dot{\ve{v}}_i + \zeta_i
\ve{v}_i - \ve{F}_i(\ve{r}^N) -z_i  \ve{E}  \right ]^2}{4 \zeta_i}  \,   
\end{align}
where the stochastic calculus is interpreted in the It\^o sense. We will
consider trajectories in the limit that $\tN$ is large so that only time extensive quantities are relevant. 

The addition of a perturbing field on the system adds an extra drift to the Gaussian action. As a consequence, we can write the ratio of the probability to observe a trajectory in the presence of the field, $\ve{E}=\ve{E}_r+\Delta \ve{E}$, relative to the probability to observe a trajectory with $\ve{E}_r$, 
\begin{equation}
\label{Eq:Measures2}
\frac{P_{\ve{E}_r+\Delta \ve{E}}[\ve{X}(\tN)]}{P_{\ve{E}_r}[\ve{X}(\tN)]} = e^{\beta \Delta U_{\Delta \ve{E}}[\ve{X}(\tN)]}
\end{equation}
where the dimensionless relative action, $\beta \Delta U_{\Delta \ve{E}}[\ve{X}(\tN)]$, can be expressed compactly as a sum of three terms, depending on their symmetry under time reversal,\cite{gao2018nonlinear}
\begin{align}
\label{Eq:DeltaUDeltaE}
\frac{\Delta U_{\Delta \ve{E}}}{\tN} =  \left [J  +  Q - E_r \sigmaNE V \right ]
\frac{\Delta E}{2} - \sigmaNE V \frac{\Delta E^2}{4}  \, 
\end{align}
 where for simplicity we take the field along one cartesian direction so that the relative action depends only on its magnitude.
The first term is asymmetric under time reversal and identified as the excess
entropy production due to the increased nonequilibrium driving. It is given by
the product of the total, time averaged ionic current in the direction of the field, 
\begin{equation}
\label{Eq:Current}
J[\ve{X}(\tN)] = \frac{1}{\tN}  \int_0^{\tN} dt \, j(t) \, ,\quad j(t)= \sum_{i=1}^N z_i v_i(t) 
\end{equation}
and the extra field $\Delta E /2$. The second term in Eq.~Ê\ref{Eq:DeltaUDeltaE} is symmetric under time reversal and referred to as the excess frenesy,\cite{basu2015nonequilibrium} where
\begin{align}
\label{Eq:Fenesy}
Q[\ve{X}(\tN)] &= \frac{1}{\tN}  \int_0^{\tN} dt \, q(t)  \; , \\
q(t)&= \sum_{i=1}^N \frac{z_i}{\zeta_i} \left [ m_i\dot{v}_i(t)- F_i(\ve{r}^N)\right ] \nonumber
\end{align}
includes the total time integrated force in the direction of the field weighted by $z_i/\zeta_i$, and a boundary term resulting in a difference in velocities at times 0 and $\tN$, times the extra field. The remaining terms are trajectory independent constants, which are proportional to
\begin{eqnarray}
\label{Eq:Sig0}
\sigmaNE  &=  \displaystyle\frac{N_c z_c^2 D_c +N_a z_a^2 D_a}{ V\kB T}
\end{eqnarray}
which is the Nernst-Einstein conductivity of the solution, where $D_i = \kB T
/\zeta_i$ is the diffusion coefficient for an isolated ion of type $i$. This decomposition of the relative action admits particularly simple, physically transparent, nonlinear response relations.\cite{baiesi2009fluctuations}

With the relative measure between trajectory ensembles defined in Eq.~\ref{Eq:Measures2}, we can follow previous work by Gao and Limmer,\cite{gao2018nonlinear} and relate nonequilibrium trajectory averages in the presence of the field, to equilibrium trajectory averages in the absence of the field. We do this by setting the reference field, $E_r=0$, so that $E=\Delta E$. For a trajectory observable, $O[\ve{X}(\tN)]$, this relation is
\begin{equation}
\label{Eq:ReweightO}
\left \langle O \right \rangle_E =\left \langle O e^{\beta \Delta U_{\ve{E}}[\ve{X}(\tN)]} \right \rangle_0
\end{equation}
where trajectory averages over the measure in Eq.~\ref{Eq:Measures}, with field value $E$, are denoted $\langle \dots \rangle_E$. Setting $O$ to 1, we find a sum rule inherited from the underlying Gaussian process that is quadratic in the field,
\begin{equation}
\left \langle e^{\beta \tN (J[\ve{X}(\tN)] + Q[\ve{X}(\tN)] )E /2} \right
\rangle_0 = e^{\beta \tN  \sigmaNE V E^2/4}
\end{equation}
which is interpretable as the ratio of nonequilibrium to equilibrium trajectory partition functions.

Identifying the joint probability of observing a value of the current and frenesy as $p_E(J,Q)=\langle \delta(J-J[\ve{X}(\tN)],Q-Q[\ve{X}(\tN)]) \rangle_E$, we can relate $p_E(J,Q)$ to its equilibrium counterpart, using Eq.~\ref{Eq:ReweightO},
\begin{equation}
\label{Eq:ReweightP}
\frac{\ln p_0(J,Q)}{\tN} = \frac{\ln p_E(J,Q)}{\tN} - \beta (J +Q) \frac{E}{2}
+ \beta \sigmaNE V \frac{E^2}{4}
\end{equation}
where we find that the nonequilbrium driving acts to reweight the joint
distribution linearly in $J+Q$, demonstrating a thermodynamic-like
relationship between this sum and its conjugate quantity $E$. Note that such
linearity is not in general valid for the marginal distribution of just the
current, $p_0(J) = \int dQ \, p_0(J,Q)$, due to correlations between $J$ and
$Q$. Equation~\ref{Eq:ReweightP} provides a route to numerically probe the tails
of nonequilibrium probability distributions using generalizations of histogram
reweighing techniques, such as those developed for equilibrium systems, like multicanonical sampling.\cite{frenkel2001understanding}

\begin{figure}[t]
\begin{center}
\includegraphics[width=8.5cm]{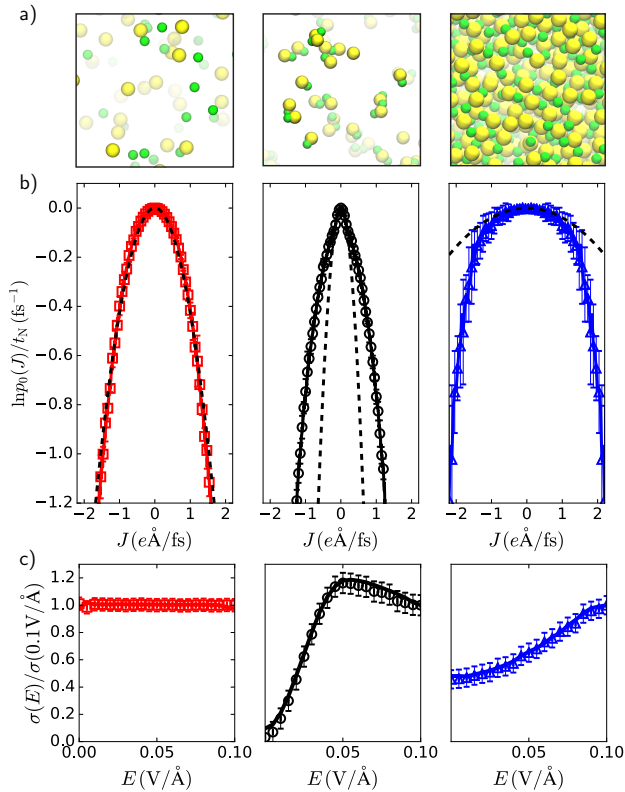}
\caption{Fluctuations and response for the 0.1 M solution of NaCl with $\epsilon_s=78.5$ (left) and 
 $\epsilon_s=10$ (middle), and the 25.3 M molten NaCl with particle number $N$. a) Characteristic snapshots of the NaCl systems considered, with green  and yellow spheres representing Na$^+$ and Cl$^-$ respectively. b)
Scaled log-probability
of the time integrated current as computed from histogram reweighting. Errorbars are one standard
deviation of the mean as computed from bootstrapping analysis. The dashed lines represent Gaussian distributions with the same mean and variance. c) Field dependent ionic conductivities relative to their values at $E=0.1$ V/$\mathrm{\AA}$. Lines are
computed from reweighting $p(J,Q)$, and symbols are computed from finite
differences of $\langle J \rangle_E$ versus $E$. Errorbars are one standard
deviation of the mean.}
\label{Fi:1}
\end{center} 
\end{figure}

With access to the joint distribution, $p_0(J,Q)$, we can compute the
relationship between the mean current and the applied electric field arbitrarily
far from equilibrium, as encoded in the electric field dependent conductivity,
$\sigma (E) = (d\langle J \rangle_E/dE)/V  $. Using Eq.~\ref{Eq:ReweightO} to first write the average current density, and then differentiating with respect to the field, we find
\begin{align}
\label{Eq:SigE}
\sigma (E) &= \lim_{\tN \rightarrow \infty} \frac{\beta \tN}{2V} \left \langle \left (\delta J^2+\delta J \delta Q \right )e^{\beta \Delta U_{\ve{E}}[\ve{X}]} \right \rangle_0 %\\
\end{align}
where $\delta O = O-\langle O \rangle$, demonstrating that $\sigma (E)$ is given by a sum of the variance of the current
and the current-frenesy correlations, reweighted by the factor that relates the
equilibrium average to the nonequilibrium ensemble at fixed $E$. Near
equilibrium ($E\approx 0$), the weight $\approx1$, and fluctuations in $J$ and $Q$ are uncorrelated due to the time reversal invariance of detailed balance dynamics, $\langle \delta J \delta Q \rangle_0=0$. In this limit Eq.~\ref{Eq:SigE} reduces to a standard Einstein-Helfand relationship.\cite{zwanzig2001nonequilibrium} For small values of $E$, we can expand the weight, and the first non-vanishing term emerges at second order in the field\cite{gao2018nonlinear}
and vanishes for uncorrelated Gaussian random variables.\footnote{{The conductivity to second order in the field is
$\frac{ \sigma(E)}{\sigma(0)} \approx 1 +  \frac{\beta^2 \tN^2 E^2 }{8 \langle J^2 \rangle_0} \left (\left\langle J^4\right\rangle_0  -  3\left\langle J^2\right\rangle_0^2 +3 \left\langle J^2 Q^2\right\rangle_0 - 3 \left\langle J^2\right\rangle_0\left\langle Q^2\right\rangle_0  \right ) $.}}

We have used these formal relationships to study the electric field dependent
conductivity of a number of different model systems.
Specifically, we have studied simple electrolytes in implicit solvent with dielectric constants
$\epsilon_s =10$ and 78.5 at $T$=300 K, and a molten salt with $\epsilon_s =1$ at $T$=1200 K. 
The main text presents results for NaCl at concentrations of 0.1 M, and 25.3 M corresponding to a molten salt. These are illustrated in Fig.~\ref{Fi:1}a), while the SM presents additional results for NaCl and MgCl$_2$ at 1.0 M. Frictions are taken to be $\zeta_i =m_i/\tau_i$ with $\tau_c=0.11$ ps for the cations and $\tau_a=0.25$ ps for the
anions, and we find $\tN =0.2$ ps is sufficient to converge the conductivity for the electrolyte system, while $\tN =0.05$ ps is sufficient for the molten salt with $\tau_c=\tau_a=1.2$ ps.\cite{SM} 
For the electrolyte solution the system size corresponds to $N_c = 100$, and for the molten salt $N_c=2500$. Long-ranged electrostatic interactions are computed using Ewald summation, and all simulations are performed with the
LAMMPS package.\cite{plimpton1995fast} 

Shown in Fig.~\ref{Fi:1}b) are the current distributions computed from
nonequilibrium molecular dynamics simulations for $E$ between 0 and 0.1~V/\AA\,
in steps of 0.01~V/\AA \, combined using Eq.~\ref{Eq:ReweightP}, followed by marginalization over $Q$. We find that for the dilute solution of NaCl with $\epsilon_s =78.5$, the current distribution is Gaussian. The Gaussian statistics follow from the largely dissociated nature of the strong electrolyte in the polar, implicit solvent, which enables ions to move free of correlations from their surrounding environment. Analogous Gaussian fluctuations are found for 1 M NaCl and MgCl$_2$ with $\epsilon_s =78.5$.\cite{SM} This is in contrast to calculations with $\epsilon_s =10$,  where ionic correlations depress motions, leading to smaller characteristic current fluctuations near equilibrium, as computed by its variance, $\tN \langle J^2 \rangle_0 $. Weaker electrolyte systems exhibit marked deviations from Gaussian statistics with enhanced probability at large values of $J$. Similar behavior is found for 1 M NaCl and MgCl$_2$ with $\epsilon_s =10$.\cite{SM} The molten salt also exhibits deviations from Gaussian statistics but with narrow tails, signifying that fluctuations are much rarer than would be expected from its large variance. The narrow distribution reflects the packing constraints that inhibit large currents.

Shown in Fig.~\ref{Fi:1}c) are the conductivities computed from $p_0(J,Q)$
continuously as a function of the applied field.
We have additionally computed the conductivity from a
numerical derivative of the average current versus applied field and find
quantitative agreement between both estimates, although the statistical errors are much larger from the finite difference approach at fixed computational cost. For strong electrolytes that
exhibit Gaussian current fluctuations, we find a field-independent conductivity,
while both the weak electrolytes and molten salt that exhibit non-Gaussian current fluctuations have
conductivities that increase with applied field. The increase is initially
quadratic, as observed experimentally\cite{harned1959physical} for dilute solutions and necessitated
by time-reversal symmetry, and plateaus at large fields. For the dilute solution the conductivity plateaus to the same value as the
strong electrolytes. The plateau value is identified as the limit of
uncorrelated motion, given by $\sigmaNE$ in Eq.~\ref{Eq:Sig0}. At intermediate fields, the dilute solution exhibits a slight maxima in conductivity as has been noted in colloids\cite{dzubiella2002lane} and low dimensional systems.\cite{kavokine2019ionic} The molten salt conductivity also increases and plateaus, though its plateau value is far below $\sigmaNE$. This field dependence of the conductivity in Fig.~\ref{Fi:1}b) is
phenomenologically known as the Onsager-Wien effect in the dilute limit.\cite{onsager1957wien,wilson1936theory} 

\begin{figure}[t]
\begin{center}
\includegraphics[width=8.5cm]{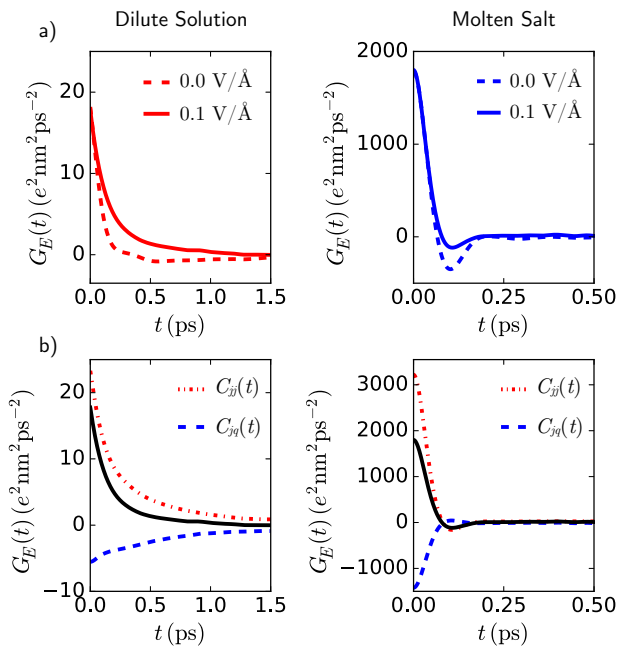}
\caption{
Time correlation function $G_E(t)=C_{jj}(t) + C_{jq}(t)$ for the field dependent conductivity of
0.1 M NaCl with $\epsilon_s=10$ (left) and molten salt (right). a) Correlations functions  for $E=0$ and
$E=0.1 \, $V$/\mathrm{\AA}$. b) Decomposition of $G_E(t)$ into current-current $C_{jj}(t)$ and
current-frenesy $C_{jq}(t)$ correlations at $E=0.1 \, $V$/\mathrm{\AA}$. The solid lines in b) are the total correlation functions.}
\label{Fi:2}
\end{center} 
\end{figure}

\begin{figure*}[th]
\begin{center}
\includegraphics[width=17cm]{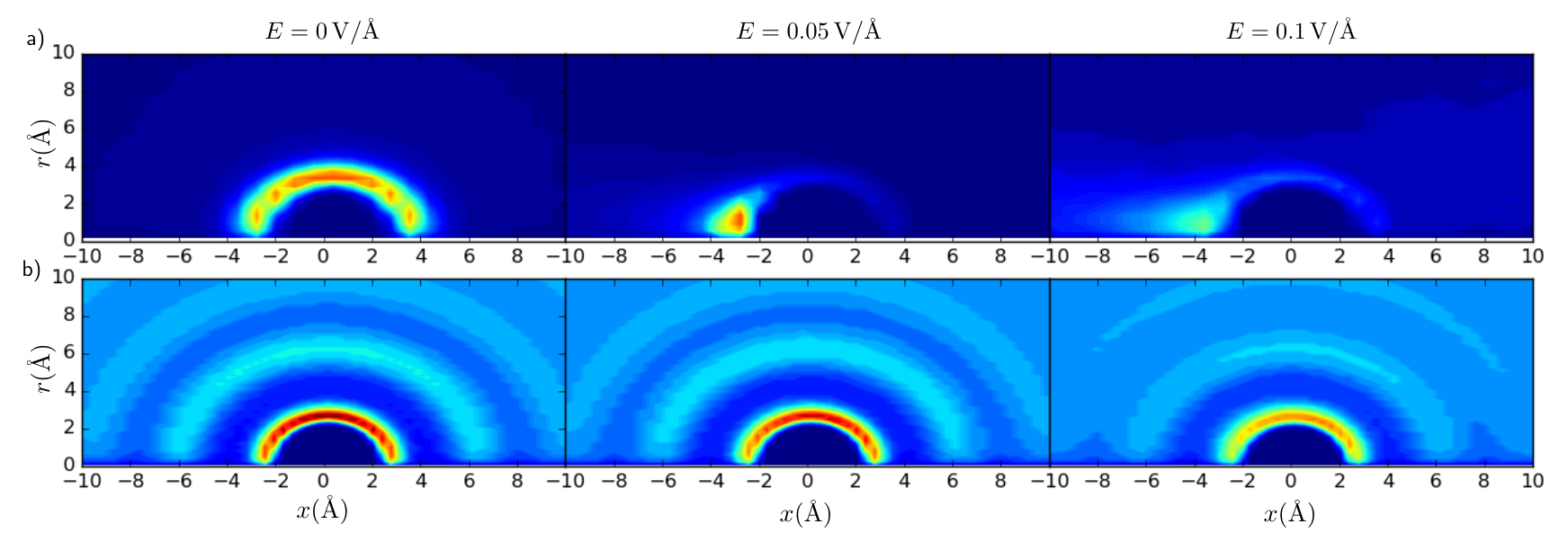}
\caption{Pair distribution functions, $g_{a,c}(\ve{r}|E)$, between Cl$^-$ and Na$^+$ with increasing field, plotted in the cylindrical coordinates for the a) dilute solution 0.1 M, $\epsilon_s=10$ and b) molten salt.}
\label{Fi:3}
\end{center} 
\end{figure*}

%The mechanism resulting in the
%nonlinearity was identified and subsequently explored as a sum of contributions
%from electrophoretic effects and ionic cloud relaxational effects, both of which
%depress the conductivity below its uncorrelated
%value.\cite{wolynes1980dynamics,demery2016conductivity,peraud2017fluctuation} At
%large applied fields, Onsager and Kim showed that the ionic correlations are
%suppressed resulting in the rise of the conductivity. 

In order to understand the effect, we can unpack
the relevant correlations using a generalized fluctuation dissipation relationship. Specifically, we rewrite the field-dependent conductivity as an average within a nonequilibrium steady-state, using
the same procedure by which we arrived at Eq.~\ref{Eq:SigE}, only now within a
trajectory ensemble at fixed $E$. In this case,
the differential response of the current an applied field is
\begin{align}
\label{Eq:SigEa}
\sigma (E) 
&=\lim_{\tN \rightarrow \infty}  \frac{\beta \tN}{2V} \left \langle \left ( \delta J^2+\delta J \delta Q \right ) \right \rangle_E \\
&=\frac{\beta}{ V}   \int_{0}^{\infty} dt\, G_E(t) \nonumber
\end{align}
where $G_E(t) = C_{jj}(t) + C_{jq}(t)$, 
$C_{jj}(t)=\langle \delta j(0) \delta j(t) \rangle_E $, 
$C_{jq}(t)= \frac{1}{2}\langle \delta j(0) \delta q(t) + \delta j(0) \delta q(-t)
\rangle_E$. The conductivity away from equilibrium is a sum of
the integrated microscopic current-current correlation function and the
integrated microscopic current-frenesy correlation function. To arrive at this
expression, we have assumed that the correlation functions decay faster than
$1/t$, and invoked the time reversal properties of $j$ and $q$ together with the
stationarity of the nonequilibrium steady-state, to eliminate one of the time
integrals. 

Fig.~\ref{Fi:2}a) shows the total time correlation
function for the conductivity, 
for both the dilute electrolyte and the molten salt with and without an applied field. In the absence of an applied field, the only nonvanishing
contribution to $G_E(t)$ is the current-current correlation function. For both
dilute electrolyte and the molten salt, current correlations decay within 1 ps. The molten salt exhibits
noticeable recoil effects evident in transient negative correlations at
intermediate times. The dilute electrolyte exhibits transient negative correlations at
intermediate times as well, though these are spread over a broader range of timescales. In both cases, the negative correlations are ionic relaxation effects that
result from ion displacements that transiently distort the local electrostatic
environment and generate a restoring force on the displaced ion from the
compensating ionic cloud left behind.\cite{wolynes1980dynamics,demery2016conductivity,peraud2017fluctuation} At high fields, this negative correlation
is suppressed, resulting in a larger integrated value of the correlation
function, hence larger conductivity. While the time-correlation functions in principle depend on the frictions in the Langevin thermostat, for the small values employed in this study, the current is independent of the friction and the frenesy depends on the friction only though the explicit factors of $\zeta_i$ in Eq.~\ref{Eq:Fenesy}.\cite{SM} 

In Fig.~\ref{Fi:2}b) we show the decomposition of $G_E(t)$ into the
current-current correlation function and the current-frenesy correlation
function for $E=0.1 \, \mathrm{V/\AA}$. For both systems, the current-current
correlation function decays slower at high fields than at $E=0$ and accounts
for the largest contribution to the $G_E(t)$ integrand. For the dilute electrolyte, positive contributions to the current-current
correlation function between unlike charges give rise to the shallow maximum at intermediate fields. These correlations are expected to be quenched out by momentum transfer to an explicit solvent.
The current-frenesy correlations are negative and the dominant contribution to the frenesy is the total charge weighted force, 
which directly manifests the depression of the conductivity due to ionic correlations.  The magnitude of the correlations in the molten salt are larger than the dilute electrolyte, reflecting its size extensive definition. The decay in the correlation function for the molten salt is nearly ten times faster, manifesting the dense system where the mean free path of a given ion is small.
For higher dielectric constant systems, the current-frenesy correlations are negligible, signifying the lack of ion correlations, and as a result, time correlation functions are independent of fields.

In the dilute solution limit, Onsager provided a theory for the field-dependent conductivity that relies on approximating the distortion of the pair correlation functions in the presence of an applied field.\cite{onsager1957wien}
In order to understand the structural origins of these dynamical effects and make contact with the previous work by Onsager, we can relate the field dependent conductivity to the change in the static ion correlations. By noting that within the steady state, the ions are force free on average, $ \langle \dot{v}_i \rangle_E=0$, we can rearrange the equation of motion in Eq.~\ref{Eq:Lang} and insert it into Eq.~\ref{Eq:Current}. This yields the average current density in the direction of the field,
\begin{align}
\label{Eq:JE}
\frac{\langle J \rangle_E}{V} = \sigmaNE E  + \sum_{i=c,a} \frac{N_i}{V} \frac{z_i}{\zeta_i} \left  \langle F_i \left (\ve{r}^N \right) \right \rangle_E
\end{align}
which is given by a sum of the Nernst-Einstein conductivity times the applied
field, and a correlated contribution from the sum of the average force acting on
ions weighted by their charge. We can express the average force in the direction
of the finite field, with unit vector $\hat{\ve{x}}$, as
\begin{align}
\label{Eq:FOW}
\left \langle F_i \left (\ve{r}^N \right) \right \rangle_E = \sum_{j=c,a}  \int d \ve{r} \, \rho_{j} g_{i,j}(\ve{r} |E) \, \hat{\ve{x}} \cdot \ve{F}^{(2)}_{i,j}(\ve{r}) 
\end{align}
where $\rho_{j}$ is the number density of the $j$th ion type, and we have introduced the pair distribution functions $g_{i,j}(\ve{r}|E)$ and the pairwise decomposable force, $\ve{F}^{(2)}_{i,j}$, between ions of type $i$ and $j$. The pair distribution function is defined as an average within the nonequilibrium steady-state
\begin{align}
 g_{i,j}(\ve{r} | E) = \frac{1}{\rho_i \rho_{j}}
\left \langle  \sum_{k \in N_i ,  l \in N_j} \delta (\ve{r}_k)  \delta (\ve{r}-\ve{r}_l)  \right \rangle_E \end{align}
normalized by the product of the densities of $i$ and $j$. In the original Onsager treatment, Eq.~\ref{Eq:JE} is assumed to have the form, $\langle J \rangle_E/V = [\sigmaNE + \Delta \sigma(E) ]E$, where $\Delta \sigma(E)$ is the correlated contribution to the conductivity computable from the knowledge of how the pair distribution function changes with applied field. 

Shown in Fig.~\ref{Fi:3}a) are the pair distribution functions between Na$^+$ and Cl$^-$ for 0.1 M and $\epsilon_s=10$, and in Fig.~\ref{Fi:3}b) for the molten salt, as a function of increasing applied field. In the presence of the field, the correlations deviate from spherical symmetry. As a consequence, we plot $g_{i,j}(\ve{r} | E)$ as a function of distance in direction of the applied field, $x$, and orthogonal radial coordinate, $r$, as the correlations do retain cylindrical symmetry. With increasing field, the correlations are found to distort away from spherical symmetry, polarizing in the direction of the applied field. This is more evident in the dilute solution compared to the molten salt. At large applied fields, the amplitude of the correlations decrease dramatically for the dilute solution, clarifying the limit of uncorrelated motion noted in Fig.~\ref{Fi:1}b). Within the molten salt, correlations persist, as even large fields are insufficient to mitigate packing constraints.

In conclusion, we have leveraged recent developments in the theory of nonequilibrium systems to relate ionic conductivities to microscopic correlations under arbitrarily large electric fields. We have found that both the fluctuations of the ion's displacement as well as the dynamical fluctuations of the intrinsic electric fields acting on an ion, affect the response of the ionic current to an additional external field. This additional contribution is absent in the Green-Kubo expression for the conductivity near-equilibrium.\cite{zwanzig2001nonequilibrium} 
This approach of reweighting nonequilibrium trajectories is general, and we expect will find use more broadly in other cases of molecular transport.
It will be particularly interesting to apply these new statistical tools to investigate the nonlinear response of ionic
liquids~\cite{daily_ionic_2009,heid_solvation_2018}, as well as transport near charged interfaces, such as nonlinear electrofriction on
corrugated surfaces\cite{Netz2003ChargedSurfaces} or nonlinear electro-osmotic response.  
 
\emph{Acknowledgments} 
The authors thank Lyd\'eric Bocquet for useful discussions.
%\todor{Do you want to thank other colleagues?}
DL and BR acknowledge financial support from the 
French Agence Nationale de la Recherche (ANR) under Grant No. ANR-17-CE09-0046-02
(NEPTUNE). CYG was supported by the U.S. Department of Energy, Office of Basic Energy Sciences
through Award Number DE-SC0019375.  BR and DTL were supported by the
France-Berkeley Fund from the University of California, Berkeley. 
\end{document}